\documentclass[aps,prl,twocolumn,superscriptaddress,10pt]{revtex4-2}
\usepackage{graphicx}
\usepackage[compat=1.1.0]{tikz-feynman}
\usepackage{amsmath, amsfonts, amssymb, bm}
\usepackage{float}
\usepackage{slashed}
\usepackage{calrsfs}
\usepackage{hyperref}
\usepackage{cleveref}
\usepackage{appendix}
\usepackage{mathrsfs}
\usepackage{enumitem}
\usepackage{lineno}

\DeclareMathAlphabet\mathbfcal{OMS}{cmsy}{b}{n}

\def \[#1\]{\begin{equation}\begin{split}
#1
\end{split}\end{equation}
}

\newcommand{\barr}[1]{\overset{}{argument}  }

\DeclareMathAlphabet{\pazocal}{OMS}{zplm}{m}{n}
\DeclareMathAlphabet\mathbfcal{OMS}{zplm}{b}{n}

\newcommand{\blue}[1]{\textcolor{black}{#1}}

\definecolor{auburn}{rgb}{0.43, 0.21, 0.1}
\definecolor{burgundy}{rgb}{0.5, 0.0, 0.13}
\definecolor{burntorange}{rgb}{0.8, 0.33, 0.0}
\definecolor{amethyst}{rgb}{0.6, 0.4, 0.8}
\definecolor{darkcerulean}{rgb}{0.03, 0.27, 0.49}
\definecolor{applegreen}{rgb}{0.55, 0.71, 0.0}
\definecolor{chocolate}{rgb}{0.82, 0.41, 0.12}
\definecolor{darkgreen}{rgb}{0.0, 0.5, 0.0}
\definecolor{violet}{rgb}{0.54, 0.17, 0.89}
\definecolor{azzurro}{rgb}{0.0, 0.44, 1.0}

\begin{document}

\title{Exact solution of the DeWitt-Brehme-Hobbs equation in copropagating electromagnetic and gravitational waves }
\author{Giulio Audagnotto}
\email{giulio.audagnotto@unito.it}
\affiliation{Max Planck Institute for Nuclear Physics, 69117 Heidelberg, Germany}
\affiliation{Department of Physics and Astronomy, University of Rochester, Rochester, NY 14627, USA}
\affiliation{Dipartimento di Fisica, Universit\`a di Torino \& INFN, Sezione di Torino, \\
Via Pietro Giuria 1, I-10125 Turin, Italy}
\author{Antonino~Di Piazza}
\email{a.dipiazza@rochester.edu}
\affiliation{Department of Physics and Astronomy, University of Rochester, Rochester, NY 14627, USA}
\affiliation{Laboratory for Laser Energetics, University of Rochester, Rochester, NY 14623, USA}

\begin{abstract}
%

%
An accelerated charge interacts with \blue{its own electromagnetic field}, a phenomenon known as electromagnetic radiation reaction. The DeWitt-Brehme-Hobbs (DWBH) equation describes the motion of a charged mass in the presence of combined electromagnetic and gravitational fields, taking into account electromagnetic radiation-reaction effects. Here, we find the first exact analytical solution of the DWBH equation in the case of a charged mass in the presence of copropagating and otherwise arbitrary electromagnetic and gravitational plane waves. 
\blue{As a consequence of the Penrose limit, the scenario considered here can be seen as a local limit around ultrarelativistic trajectories in a general curved spacetime.}
Finally, the paradigmatic example of an electromagnetic wave in the presence of a constant-amplitude gravitational wave is worked out explicitly and it is shown how the presence of the gravitational wave can qualitatively change electromagnetic radiation-reaction effects.
\end{abstract}

\maketitle
\paragraph{Introduction.}
When a charged particle is accelerated, the electromagnetic field that it produces undergoes a deformation. For this reason, the study of the motion of an electrically charged particle in general cannot ignore the dynamics of its electromagnetic field and the consequent interaction between the charge and the field itself \cite{Jackson_b_1975,Landau_b_2_1975}. In flat spacetime, this led to a modification of the Lorentz equation due to Dirac, Abraham, and Lorentz himself \cite{Lorentz_b_1909, Abraham_b_1905, Dirac_1938}. The additional force resulting from the interaction between the charge and its field is known as radiation-reaction force and it has puzzled scientists for decades. The complete equation of motion, known as the Lorentz-Abraham-Dirac (LAD) equation, in fact, is not Newtonian and contains a term involving the time-derivative of the acceleration. Such an equation is inevitably pathological and admits unphysical solutions. However, a reduction of order, first proposed by Landau and Lifshitz \cite{Landau_b_2_1975}, can be applied to the LAD equation to obtain an equation known as the Landau-Lifshitz (LL) equation, which does not feature any of the shortcomings of the LAD equation while admitting its physical solutions \cite{Spohn_2000}. In the case of an electron, it can be proved that the LL equation is classically equivalent to the LAD equation, in the sense that their differences are smaller than the already ignored quantum effects \cite{Landau_b_2_1975,Rohrlich_b_2007}. 

The picture becomes further complicated when one includes gravitational effects. Indeed, in curved spacetimes light can propagate off the light-cone, such that the electromagnetic radiation emitted by a massive charge can interact again with the charge itself after a finite propagation time \cite{Poisson:2011nh}. In other words, the Huygens' principle in general does not hold. This effect has been taken into account in the work of DeWitt, Brehme, and Hobbs \cite{DeWitt:1960fc, Hobbs:1968}, who generalized the LAD equation to curved spacetime. The main novelty of this equation is the appearance of a ``tail'' term involving the portion of the electromagnetic Green's function supported inside the light-cone. This exactly describes the aforementioned failure of the Huygens' principle. As in flat spacetime, it is possible to operate a reduction of order to the LAD terms in the equation, thereby fixing the pathologies arising from the third-order derivative terms \cite{Quinn:1996am}.
Below, we will refer to the order-reduced equation as to the DWBH equation.

Due to its mathematical complexity, it is challenging to solve the DWBH equation analytically and, to the best of our knowledge, no analytical solution of this equation has been found as of today. This is hardly surprising, given that only a handful of solutions of the LL equation in flat spacetime are known. A relevant example here is the solution found in a plane-wave background field \cite{Di_Piazza_2008_a}. Despite featuring idealized highly-symmetric structures, plane waves are extremely useful models. Indeed, any wavefront looks approximately locally planar and an arbitrary electromagnetic field looks approximately like a plane wave in the rest frame of an ultrarelativistic charge \cite{Jackson_b_1975}.
Moreover, plane waves can also be employed to efficiently describe lasers fields, especially if they are not tightly focused \cite{Di_Piazza_2012,Gonoskov_2022,Fedotov_2023}. Notably, the same is true for gravitational plane waves: An arbitrary spacetime locally resembles a plane-wave spacetime for an ultrarelativistic observer, which is known as Penrose limit \cite{Penrose1976}.

In the present Letter we find the exact analytical solution of the DWBH equation in the case of an electromagnetic plane wave propagating in a generic gravitational plane wave spacetime along the same direction and otherwise arbitrary. \blue{By constructing the vector Green's function from the solutions of the wave equation, we show that the tail term of the DWBH equation identically vanishes in plane-wave spacetimes. The same conclusion has been drawn through different derivations in Ref. \cite{Harte:2013dba} for Ricci-flat plane waves and, in general, in Ref. \cite{Kunzle1968}.} \blue{Although the scenario considered here can be viewed as the general local limit around ultrarelativistic observers according to Penrose limit, the resulting particle trajectory may not provide an accurate description of the actual motion in a generic metric. In fact, radiation–reaction effects are inherently non-local phenomena. Therefore, the absence of tail terms in the present scenario implies that, in general, such nonlocal effects must arise from higher-order terms in the expansion of the null Fermi coordinates. Resulting accumulation effects may also strongly influence the particle's dynamics.}

\paragraph{Electromagnetic radiation reaction.} 
As previously mentioned, the study of the electromagnetic self-force has a very long history (see the books \cite{Jackson_b_1975,Landau_b_2_1975,Barut_b_1980,Rohrlich_b_2007} and the reviews \cite{Di_Piazza_2012,Burton_2014,Gonoskov_2022,Fedotov_2023}). 
Here, we directly report the LL equation \cite{Landau_b_2_1975}
\begin{align}
\label{LL eq.}
\frac{d u^\alpha}{d\tau} =\ 
& \frac{e}{m} F^{\alpha \beta} u_\beta + \tau_e\frac{e}{m} \bigg(
\frac{d F^{\alpha \beta}}{d\tau} u_\beta \nonumber \\
& 
+ \frac{e}{m} F^{\alpha \beta} F_\beta^{\ \gamma} u_\gamma 
- \frac{e}{m} u_\delta F^{\delta \beta} F_\beta^{\ \gamma} u_\gamma u^\alpha
\bigg),
\end{align}
where $e$ and $m$ are the charge and mass of the electron, respectively, $\tau$ is its proper time, $u^{\mu}$ its four-velocity, $F^{\mu\nu}(x)$ the external electromagnetic field, and $\tau_e = e^2/6\pi m$ (units with $\epsilon_0=c=1$ are used throughout and the flat metric tensor $\eta_{\mu\nu}$ is $\text{diag}(+1,-1,-1,-1)$).
It is worth mentioning that the LL equation has also been derived in Ref. \cite{Gralla_2009} using a non-pointlike description of the charge and thus rigorously avoiding Coulomb-like divergences. 

In Ref.~\cite{DeWitt:1960fc} DeWitt and Brehme extended the LAD equation to curved spacetime obtaining a close but incomplete equation, which was further corrected by Hobbs \cite{Hobbs:1968}. The resulting full equation of motion was also derived later in Ref. \cite{Quinn:1996am}, where it is also shown that the pathologies of the LAD equation in a curved spacetime can be cured via a reduction of order resulting in a general covariant version of the LL equation (see also Ref. \cite{Gralla_2008}). This order-reduced equation, which we refer to as the DWBH equation, reads \cite{Quinn:1996am}
\begin{align}
\label{DWBH eq.}
&\frac{D u^\alpha}{D \tau} =\ 
 \frac{e}{m} F^{\alpha \beta} u_\beta + \tau_e \frac{e}{m}  \bigg(
\frac{D F^{\alpha \beta}}{D \tau} u_\beta + \frac{e}{m} F^{\alpha \beta} F_\beta^{\ \gamma} u_\gamma
\nonumber
\\ &\quad
- \frac{e}{m} u_\delta F^{\delta \beta} F_\beta^{\ \gamma} u_\gamma u^\alpha 
\bigg) + \frac{\tau_e}{2} \left( R^\alpha{}_\beta u^\beta - R_{\mu \nu} u^\mu u^\nu u^\alpha \right) \nonumber \\
&\quad + 3 \tau_e u_\nu \int_{-\infty}^{\tau - 0^+}
d\tau' \nabla^{[ \alpha } G_{\text{ret}}^{\nu]}{}_{\lambda'} u^{\lambda'}(\tau'),
\end{align}
where $D/D\tau$ is the covariant derivative along the trajectory, $a^{[ \alpha}b^{\beta ]}=(a^{\alpha}b^{\beta}-a^{\beta}b^{\alpha})/2$ for two arbitrary four-vectors $a^{\mu}$ and $b^{\mu}$, $R^{\mu\nu}(x)$ is the Ricci tensor, and where the retarded Green's function $G_{\text{ret}}^{\nu}{}_{\lambda'}(x(\tau),x(\tau'))$ of the covariant electromagnetic wave equation is calculated in two spacetime points belonging to the particle worldline.

While the curved spacetime generalization of the LL equation is clearly recognizable, the last two terms of Eq. \eqref{DWBH eq.} are genuine gravitational features.
The contributions involving the Ricci tensor can be seen as a modification of the incoming field due to its interaction with the spacetime curvature \cite{Hobbs:1968}. 
The last term is the aforementioned tail term, which is not local and depends on the past history of the particle \cite{Poisson:2011nh}.
\paragraph{Green's function in plane-wave spacetimes.}
The metric describing a plane-wave spacetime propagating along the direction $\bm{n}$ can be written as \cite{Einstein:1937qu,Bondi:1957,Bondi:1958aj}
\[\label{Rosen metric}
g_{\mu\nu}(\phi) = n_{\mu} \tilde{n}_{\nu}+n_{\nu} \tilde{n}_{\mu} + \gamma_{ij}(\phi)\delta^i_\mu  \delta ^j_\nu,
\]
where $n^{\mu}=(1,\bm{n})$, $\tilde{n}^{\mu}=(1,-\bm{n})/2$, while $i,j$ refer to the two coordinates transverse to the wave propagation $\bm{n}=\bm{z}$ and $\phi \equiv x^- = n\cdot x = t-z$.
The last coordinate $x^+=\tilde{n}\cdot x$ completes the light-cone set $\{x^-, x^i, x^+   \}$.
The metric tensor $g_{\mu\nu}(\phi)$ in Eq. \eqref{Rosen metric} is known as Rosen metric and, despite not being globally defined \cite{Penrose:1965rx, Harte:2012caustics}, it is extremely useful to describe the dynamics, \blue{due} to its high degree of symmetry and its dependence on a single coordinate.
Another chart, known as Brinkmann chart, is used to describe plane-wave spacetimes \cite{Brinkmann:1925fr}, which has the advantage of being globally defined and the disadvantage of making the dynamics equations more complicated than in the Rosen chart. The metric in this chart has a Kerr–Schild form
$\mathcal{G}_{\mu\nu}(\phi, X^i) = \eta_{\mu\nu}
+ H_{ij}(\phi)X^iX^j n_\mu n_\nu$, where $H_{ij}(\phi)$ is the Brinkmamnn wave profile and it is connected to the Rosen metric vierbein $e^\alpha{}_{ \mu }(\phi)$, defined as $e_{\alpha \mu }(\phi) e^\alpha_{\-\ \nu}(\phi) = g_{\mu\nu}(\phi)$, through the equation $\ddot{e}_{ij}(\phi) = H_{ik}(\phi)e^k_{\-\ j}(\phi)$
(here and below the dot indicates the derivative with respect to $x^- = \phi$).
Throughout this work, we adopt the convention that the left index of a vierbein is always Minkowskian, while the right index always refers to curved spacetime and is lowered and raised by the metric tensor $g_{\mu\nu}(\phi)$.

In plane-wave spacetimes, the Huygens' principle has been shown to hold \cite{Kunzle1968,Friedlander:1975, Harte:2012caustics, Gibbons:1975jb}. Thus, the Green's function of the physical electromagnetic field $F_{\mu\nu}$ does not feature any tail term \cite{Harte:2013dba, Kunzle1968} and the last term in the DWBH equation \eqref{DWBH eq.} identically vanishes. Here, we provide a simple proof of this statement \blue{which highlights some interesting connections between particles and fields dynamics in plane waves. }

\blue{With this aim, we will first briefly introduce the main features of the particle dynamics in Rosen coordinates, and then we will exploit these notions to construct the four-vector solutions of the wave equation and the corresponding propagator.}
It can be shown that the geodesic motion in a Rosen chart can be fully described through a simple operator $\Lambda^\alpha_{ p, \beta}(x^-)$ \cite{Audagnotto:2024iel}. In fact, the vierbein-projected four-momentum $\bar{\pi}_p^\alpha(x^-)=e^{\alpha}{}_{\mu}(x^-)\pi_p^{\mu}(x^-)$ of a free-falling point-like particle with initial momentum $p^\alpha$ takes the form \[\label{momentum rosen evo}
\bar{\pi}_p^\alpha(x^-)  = \Lambda^\alpha_{ p, \beta}(x^-) p^\beta,
\]
where
\[
\Lambda_p(x^-) = \exp\left\{
- 2 \frac{p_i}{p^-}
n^{[ \alpha} \Delta e ^{\beta ] i}(x^-) 
\right\},
\]
with $\Delta e ^{\beta i}(x^-) = e^{\beta i}(x^-) -\eta^{\beta i}$.
From this property it follows that the momentum at a generic value of $x^-$ is related to its value at another $y^-$ through the expression 
\[
\bar{\pi}_p^\alpha(x^-)  = \Lambda^\alpha_{ p, \beta}(x^-)\Lambda^{\gamma' \beta}_p(y^-) \bar{\pi}_{p, \gamma'}(y^-) ,
\]
such that the bitensor 
\[ \label{parallel prop. p}
\bar{g}^{\alpha \gamma'} (x^-,y^-;p) & = \Lambda^\alpha_{ p, \beta}(x^-)\Lambda^{\gamma' \beta}_p(y^-)
\\ 
&   =
\eta^{\alpha \gamma'}
	- 2 \frac{p_i}{p^-}
	n^{[\alpha} \delta e^{\gamma'] i}(x^-,y^-)
\\&  
	- \frac{p_i p_j}{2 (p^-)^2} \delta e_{k}{}^{ i}(x^-,y^-)\delta  e^{k j}(x^-,y^-) n^\alpha n^{\gamma'},
\]
with $\delta e^{k i}(x^-,y^-) = e^{k i}(x^-) -e^{k i}(y^-)$,
is exactly the vierbein-projected parallel propagator along the geodesic with initial tangent vector $p^\alpha$ \cite{Poisson:2011nh}.

Let us now consider the massless four-vector wave equation in the Lorenz gauge
\[
\nabla_\mu \nabla^\mu  A^\nu 
+ R^\nu_{\-\ \mu}A^\mu
= 0  .
\] 
In a Rosen chart, the vierbein-projected solution of this equation $\bar{A}_{q,r}^\alpha(x) = e^\alpha{}_\mu(x^-) A_{q,r}^\mu(x) $, with initial four-momentum $q^\mu$ and polarization $r=0,\ldots,3$, can be written as \cite{Audagnotto:2024iel}
\[
\bar{{A}}_{q,r}^\alpha(x) = \Omega(x^-) e^{iS_q(x)}\bar{\mathcal{E}}_{q,r}^\alpha(x^-) ,
\] 
where $\Omega(x^-) = |\det g(x^-)|^{-\frac{1}{4}} \equiv |\det \gamma(x^-)|^{-\frac{1}{4}} $, $S_q(x)$ is the classical Hamilton-Jacobi action for a particle in a plane wave 
\[
S_q(x) &= - q\cdot x 
+ \frac{q_i q_j}{2 q^-} \int^{x^-} d\tilde{\phi}\,
[\gamma^{ij}(\tilde{\phi}) - \eta^{ij}]
\]
and $\bar{\mathcal{E}}^\alpha_{q,r}(x^-)$ is the polarization four-vector 
\[
\bar{\mathcal{E}}^\alpha_{q,r}(x^-) = 
\Lambda^\alpha_{ q, \beta}(x^-)\varepsilon_{q,r}^\beta
-
i\frac{ \varepsilon^-_{q,r}}{ 2q^-} \sigma(x^-)  n^\alpha ,
\]
with $\varepsilon_{q,r}^\beta$ being the polarization vector of a free photon with momentum $q^\mu$ and polarization $r$ in flat spacetime and $\sigma(x^-)$ the trace of the two-by-two matrix $\sigma_{ij}(x^-)=\dot{e}_{ik}(x^-)e_j{}^{k}(x^-)$. It is worth observing that $\Phi_q(x) = \Omega(x^-) e^{iS_q(x)}$ is the solution of the scalar wave equation $\nabla_\mu \nabla^\mu \Phi 
= 0   $ with the same asymptotic four-momentum, such that the four-vector solution could be concisely written in the form $\bar{{A}}_{q,r}^\alpha(x) = \Phi_q(x) \bar{\mathcal{E}}_{q,r}^\alpha(x^-) $. 

We can now introduce the vierbein-projected retarded Green's function $\bar{G}^{\alpha \beta'}_{\text{ret}}(x,y) $, satisfying the equation 
    $\bar{\nabla}_\gamma \bar{\nabla}^\gamma \bar{G}^{\alpha \beta'}_{\text{ret}}(x,y)
	+
	\bar{R}^\alpha_{\-\ \gamma}\bar{G}^{\gamma \beta'}_{\text{ret}}(x,y)
	=  |\det g(x^-)|^{-\frac{1}{2}} \bar{g}^{\alpha \beta'}(x,y) \delta^{(4)}(x-y)$, as \cite{Hollowood:2008kq}
\[
\bar{G}^{\alpha \beta'}_{\text{ret}}(x,y) &= -\int_{\text{ret}}  \frac{d^4 q}{(2 \pi)^4q^2}
\sum_r\eta_{rr} \bar{A}^{*\alpha}_{q,r}(x)\bar{A}^{\beta'}_{q,r}(y),
\]
where the index $\text{“ret”}$ indicates that the poles at $q^2=0$ have to be circumvented in order to provide the retarded Green's function and where the sum is understood to be over the flat polarizations with the convention $\sum_r\eta_{rr}\varepsilon^{* \mu}_{q,r} \varepsilon_{q,r}^\nu = \eta^{\mu\nu}$ \cite{Itzykson_b_1980}. 
A straightforward calculation shows that 
\[
\sum_r\eta_{rr}
\mathcal{\bar{E}}^{*\alpha}_{q,r}(x^-)&\mathcal{\bar{E}}^{\beta'}_{q,r}(y^-)  
= \bar{g}^{\alpha \beta' } (x^-,y^-;q)
\\& 
+ \frac{ i}{2q^-}
\left[
\sigma (x^-)
- 
\sigma (y^-)
\right]n^\alpha n^{\beta '},
\]
such that 
\[ \label{Vector Green 2}
\begin{aligned}
&\bar{G}^{\alpha \beta'}_{\text{ret}}(x,y)
= 
- \int_{\text{ret}} \frac{d^4 q}{(2\pi)^4 q^2} 
\Phi^*_q(x) \Phi_q(y)
\\
& \times \bigg\{
\bar{g}^{\alpha \beta'}(x^-, y^-; q)
+ \frac{i}{2q^-} \left[
\sigma(x^-)
- \sigma(y^-)
\right] n^\alpha n^{\beta'}
\bigg\}.
\end{aligned}
\]

If we exclude the tensorial terms inside the curly brackets, this integral reduces to the scalar Green's function, which can be easily shown to be \cite{Gibbons:1975jb,Adamo:2022qci, Harte:2012caustics}
\[
G_{\text{ret}}(x,y) = \frac{\theta(x^-  -  y^- )}{4 \pi}\sqrt{\Delta(x^-, y^-)} \delta( \Sigma(x,y)),
\]
where $\Sigma(x,y)$ is the Synge world function \cite{Synge:1960ueh} and $\Delta(x^-, y^-)$ is the van Vleck determinant \cite{Poisson:2011nh, van_Vleck_1928}. By defining
$\blue{\mathcal{F}}^{ij} (x^-,y^-) = \int_{y^-}^{x^-} d \tilde{\phi} \gamma^{ij}(\tilde{\phi}) $ and $z^\mu = x^\mu - y^\mu $, these read
\[
\begin{aligned}
&\Sigma(x, y) = 
z^-
\left[
z^+ +  \frac{1}{2 }z^i \blue{\mathcal{F}}^{-1}_{ij}(x^-,y^-) z^j
\right],
\\&
\Delta(x^-, y^-) = \frac{
(z^-)^2\Omega^2(x^-) \Omega^2(y^-)}{
\det \blue{\mathcal{F}}^{ij}(x^-,y^-)
}.
\end{aligned}
\]

Here and below we assume the points $x,y$ to be connected by a single geodesic (see \cite{Harte:2012caustics} for further details).
The scalar propagator is only supported on the light-cone, therefore the Huygens' principle holds manifestly for scalar fields in plane-wave spacetimes. 
However, when the tensorial components in Eq. \eqref{Vector Green 2} are included, off-light-cone contributions arise. A direct calculation shows that 
\[
&\bar{G}^{\alpha \beta'}_{\text{ret}}(x,y) 
=
\frac{\theta(z^-)}{4 \pi }\sqrt{\Delta(x^-, y^-)}  \bigg\{
\bar{g}^{\alpha \beta' } (x^-, y^-) \delta(\Sigma(x,y))
\\ & \quad 
- V(x^-, y^-)
n^\alpha n^{\beta'} \theta(\Sigma(x,y))
\bigg\},
\]
where $\bar{g}^{\alpha \beta' } (x^-, y^-)$ is the general vierbein-projected parallel propagator obtained replacing $ p_i/p^- $ with $  \blue{\mathcal{F}}^{-1}_{ij}(x^-,y^-)z^j $ in Eq. \eqref{parallel prop. p}, while the tail term reads
\[
&V(x^-, y^-)= 
\frac{1}{2z^-}
\bigg[
\sigma(x^-)
- \sigma(y^-)
\\& \quad 
+
\blue{\mathcal{F}}^{-1} _{ij}(x^-, y^-)  \delta e_{k}{}^{ i}(x^-, y^-) 
\delta e^{k j}(x^-, y^-) 
\bigg].
\]
By exploiting the equation $\dot{\sigma}_{ij} = H_{ij} - \sigma_{ik}\sigma^k{}_j
$ \cite{Audagnotto:2024iel}, it is easy to show that the coincidence limit of the tail term restores the Ricci tensor, as expected for the Hadamard construction \cite{Poisson:2011nh}
\[
\begin{split}
\lim_{y^- \rightarrow x^-}
&\frac{\theta(z^-)}{4 \pi }\sqrt{\Delta(x^-, y^-)} V(x^-, y^-) n_\alpha n_{\beta'}\\
&\quad=  \frac{1}{8 \pi } H(x^-) n_\alpha n_{\beta'}
=  \frac{1}{8 \pi } R_{\alpha\beta'}(x^-),
\end{split}
\]
where the limit is performed for $z^->0$ and $H(x^-)=H^i{}_i(x^-)$. The key feature of the vector Green's function is that its tail is a pure gauge contribution. In fact, the tail \blue{term multiplying the $\theta$ function} depends only on $x^- = n\cdot x$ and its tensor structure is given by $n_\alpha n_{\beta'}$, such that the Green's function for the physical electromagnetic field appearing in Eq. \eqref{DWBH eq.} features no tail. In other words, $\nabla^{[ \alpha } G_{\text{ret}}^{\nu]}{}_{\lambda'}(x,y)$ is supported only on the light-cone, the Huygens' principle is preserved\blue{, and the tail term in the DWBH equation, i.e., the last line of Eq. (\ref{DWBH eq.}) vanishes.} \cite{Harte:2013dba, Friedlander:1975}.

\paragraph{Solution of the DWBH equation.}
In the following, we solve analytically the resulting equation in Rosen coordinates and in the presence of an electromagnetic plane wave also propagating along the $z$ direction but otherwise arbitrary. This is described by the four-vector potential $\blue{a}_{\mu}(\phi)=\delta_{\mu}{}^i \blue{a}_i(\phi)$ with Maxwell tensor $\blue{f}_{\mu\nu}(\phi)=\partial_{\mu}\blue{a}_{\nu}(\phi)-\partial_{\nu}\blue{a}_{\mu}(\phi)=n_{\mu}\dot{\blue{a}}_{\nu}(\phi)-n_{\nu}\dot{\blue{a}}_{\mu}(\phi)$. 
Moreover, we include in our analysis the general Ricci-curved case in which the gravitational wave is generated by an energy-momentum $T_{\mu\nu}(\phi)=\rho(\phi) n_{\mu}n_{\nu}$ \cite{Blau:2002js}, which is not null in the interaction region.
A clarification is in order here: while this tensor can be interpreted as the energy-momentum of a null dust generating the gravitational plane wave, it should not be regarded as the energy-momentum tensor of the electromagnetic plane wave $\blue{f}^{\alpha\beta}(\phi)$ itself. In fact, although an electromagnetic plane wave features an energy-momentum tensor exactly of the form $T_{\mu \nu}(\phi) = \blue{U}(\phi) n_\mu n_\nu$, the effects of the gravitational field produced by the electromagnetic field are not taken into account by the DWBH equation \cite{Zimmerman}. Clearly, the solution below also applies to vacuum regions where both electromagnetic and gravitational waves freely propagate such that $T_{\mu\nu}(\phi)=0$ and $R_{\mu\nu}(\phi)=0$.

In the Rosen chart, one can verify that
\[ \label{eq. cov der}
\begin{aligned}
&\frac{D u^\alpha }{D \tau } = u^- g^{\alpha \nu }\dot{u}_\nu  - \frac{1}{2} n^\alpha  \dot{\gamma}_{ij}u^i u^j,
\\&
\frac{D F^{\alpha \beta}}{D \tau} 
= 
u^-
\left(
\dot{\blue{f}}^{\alpha\beta}
+ \dot{\blue{a}}_i  
\dot{\gamma}^{i [\alpha  } n^{\beta]} 
\right).
\end{aligned}
\]
Now, analogously to the pure electromagnetic case \cite{Di_Piazza_2008_a}, the component $u^-(\phi)$ can be isolated to obtain a solvable differential equation. Indeed, exploiting Eq. \eqref{eq. cov der} and recalling that $R_{\mu \nu }(\phi) = H(\phi) n_\mu n_\nu = - (\kappa^2/4) \rho(\phi) n_\mu n_\nu $, with $\kappa^2=32\pi G$, it is easy to find that the contraction of Eq. \eqref{DWBH eq.} with $n^\alpha$ gives
\[
\dot{u}^- 
=
\tau_e
\left(
\frac{e^2}{m^2}
\dot{\blue{a}}_i\dot{\blue{a}}_j \gamma^{ij}
+
\frac{\kappa^2}{8} \rho 
\right)
(u^-)^2.
\]
This equation is easily integrated and the solution can be written as $u^-(\phi) = u_0^-/ w(\phi )$, where $u_0^-$ is the initial value at the phase $\phi_0$ and
\[ \label{w RR}
&w(\phi)  = 1
\\ &
\quad -
\tau_e u_0^- \int_{\phi_0}^{\phi}d\tilde{\phi} \,
\left[
\frac{e^2}{m^2}
\dot{\blue{a}}_i(\tilde{\phi}) \dot{\blue{a}}_j(\tilde{\phi})  \gamma^{ij}(\tilde{\phi}) 
+
\frac{\kappa^2}{8} \rho(\tilde{\phi}) 
\right].
\]
Being $\gamma_{ij}(\phi)$ negative definite and the energy $\rho(\phi)$ always positive, we see here that the pure electromagnetic and the matter curvature contributions to $u^-(\phi)$ have opposite signs (see also below).

Knowing $u^-(\phi)$, the perpendicular components of the four-velocity are found to be
\[ 
u^i(\phi)   = \frac{\gamma^{ij}(\phi)}{w(\phi )}
\bigg[
u_{0,j} - \frac{e}{m} \mathcal{A}_j(\phi)
\bigg], 
\]
with $u_{0,j}$ being the initial transverse velocity and 
\[
\mathcal{A}_i(\phi) & =   
\int _{\phi_0}^\phi d\tilde{\phi} \, w(\tilde{\phi} )  \dot{\blue{a}}_i(\tilde{\phi})
+ 
\tau_e u_0^-[\dot{\blue{a}}_i (\phi) - \dot{\blue{a}}_i (\phi_0) ]
\\ &
+
\frac{1}{2}\tau_e u_0^-
\int _{\phi_0}^\phi d \tilde{\phi} 
\gamma_{ik} (\tilde{\phi}) \dot{ \gamma}^{ kj  }(\tilde{\phi}) \dot{\blue{a}}_j (\tilde{\phi})
.
\]
The last remaining component of the four-velocity can be found from the on-shell condition, such that the complete four-velocity can be cast in the form 
\[
\label{DWBH solution velocity}
& u^\mu (\phi)  =
\frac{1}{w(\phi )}
\bigg\{
g^{\mu \nu}(\phi) u_{0,\nu} - \frac{e}{m} \mathcal{A}^\mu(\phi)
\\&
    +
\frac{e}{m u^-_0}
\bigg[
\mathcal{A}^i(\phi)u_{0,i}
- \frac{e}{2 m}\mathcal{A}^i(\phi)\mathcal{A}_i(\phi)\bigg]n^\mu
\\ & 
+
\frac{1}{2 u_0^- }\bigg[ w^2(\phi) -1
- \bigg(
\gamma^{ij}(\phi) - \eta^{ij}
\bigg)u_{0,i}u_{0,j}
  \bigg]n^\mu
\bigg\},
\]
where $\mathcal{A}^\mu(\phi)=\delta^{\mu}{}_i\mathcal{A}^i(\phi)=\delta^{\mu}{}_i\gamma^{ij}(\phi)\mathcal{A}_j(\phi)$.

Finally, the particle trajectory can by obtained by integrating the equation $u^\mu (\phi)=dx^{\mu}(\tau)/d\tau=u^-(\phi)dx^{\mu}/d\phi$. It is straightforward to check that the solution in Eq. (\ref{DWBH solution velocity}) reduces to that reported in Ref. \cite{Di_Piazza_2008} in the flat-spacetime case and to the free-falling solution when the electromagnetic field is turned off \cite{Garriga:1990dp}.

The impact of the gravitational field on the charge dynamics can be qualitatively ascertained from Eq. (\ref{w RR}). There are two gravitational effects; one is directly related to the metric and is described by the two-by-two matrix $\gamma_{ij}(\phi)-\eta_{ij}$, while the other depends on the presence of matter and it is proportional to $\rho(\phi)$. The latter effect, as we have observed, has the opposite sign of the corresponding electromagnetic contribution. However, the gravitational effect described by $\gamma_{ij}(\phi)-\eta_{ij}$ can be either positive or negative.

In some cases, the integrals involved in the expressions of the functions $w(\phi)$ and $\mathcal{A}_i(\phi)$ can be taken analytically, and the peculiar effects of the gravitational plane wave on the charge dynamics can be explicitly shown. 

Let us consider the simple case of a sandwich gravitational plane wave in vacuum with a constant diagonal Brinkmann profile $H_{ij}(\phi)= H_+ \operatorname{diag}(1, -1)$ for $\phi \in (-\varPhi,\varPhi)$, with $\varPhi=\pi/(2\sqrt{H_+})$, and zero elsewhere, such that $\gamma_{ij}(\phi)=-\operatorname{diag}(\cos^2(\sqrt{H_+}\phi), \cosh^2(\sqrt{H_+}\phi))$. 
In addition we assume that \blue{$a_i(\phi) = \delta_i{}^1 a_0 \cos(\omega\phi)\cos^2(\sqrt{H_+}\phi)$, corresponding to $a^i(\phi) = -\delta^i{}_1 a_0 \cos(\omega\phi) $}. 
Passing to the dimensionless phase variable $\varphi=\omega\phi$ and introducing the parameter $\lambda=\sqrt{H_+}/\omega$, one obtains for $\varphi, \varphi_0 \in (-\omega \varPhi, \omega \varPhi)$ that (see Eq. (\ref{w RR}))
\begin{equation}
\label{w}
\begin{split}
w(\varphi)=1
 &+
\omega\tau_eu_0^-\xi_0^2
\int_{\varphi_0}^{\varphi} d\tilde{\varphi} \, \left[ 2 \lambda \sin(\lambda \tilde{\varphi}) \, \cos( \tilde{\varphi}) 
\right.
\\ &  
\left.
+  \cos(\lambda\tilde{\varphi}) \, \sin(\tilde{\varphi}) \right]^2,
\end{split}
\end{equation}
where $\xi_0=|e|\blue{a}_0/m$. The integrals in the functions $\mathcal{A}_i(\phi)$ can clearly also be taken analytically, but the resulting cumbersome expressions are not particularly illuminating.
\begin{figure}
		\centering
		\includegraphics[width=0.9\linewidth]{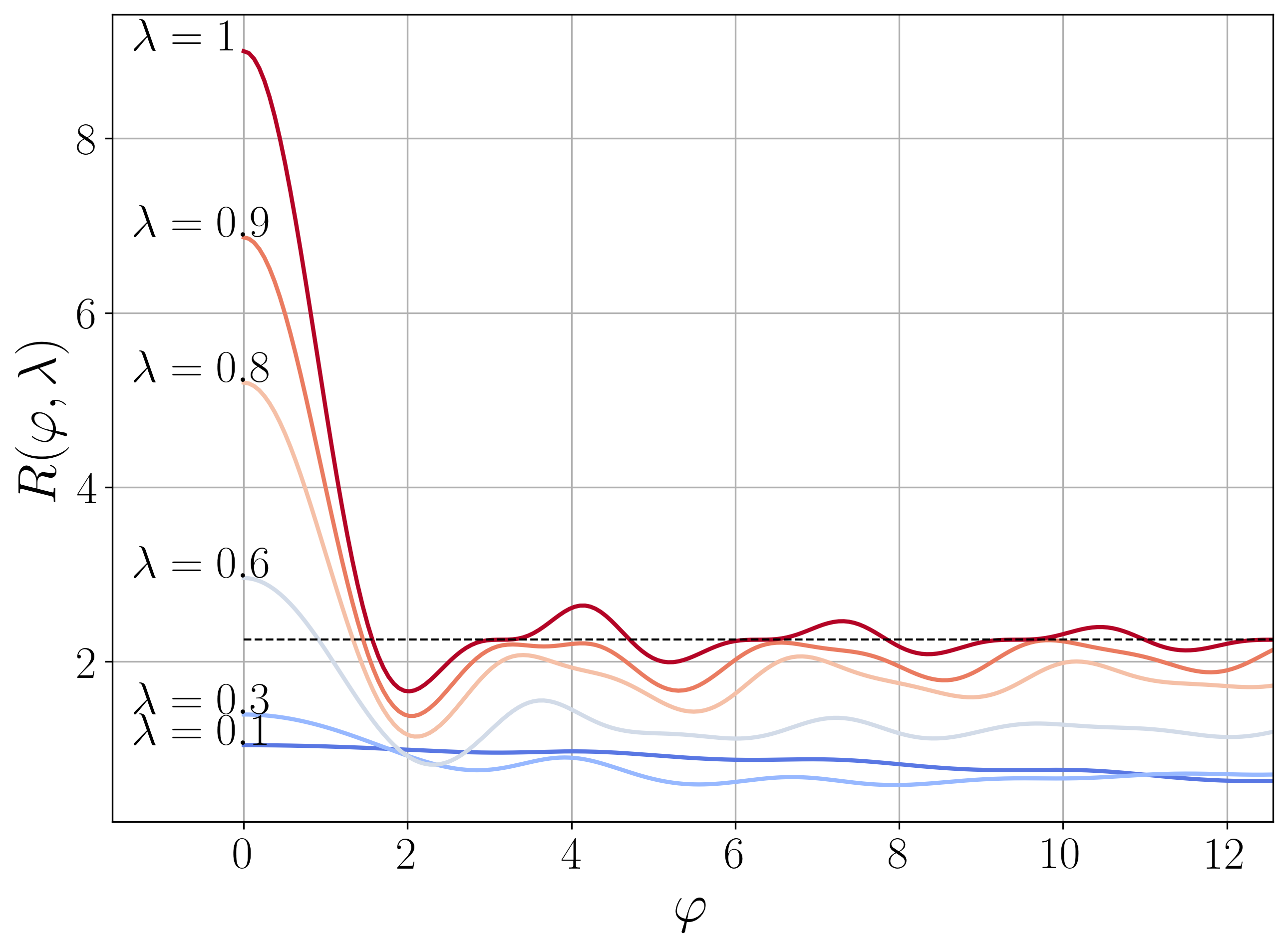}
		\caption{\blue{The function $R(\varphi, \lambda) = \frac{(w(\varphi) - 1)_{\lambda}}{(w(\varphi) - 1)_{\lambda = 0 } } $ for $a^i(\phi) = -\delta^i{}_1 a_0 \cos(\omega\phi) $ and different values of $\lambda$. The black dashed line at 9/4 represents the asymptote of $R(\varphi, 1)$.}}
		\label{fig:w_cos_cossquared}
\end{figure}
Looking at Eq. (\ref{w}) and \blue{Fig. \ref{fig:w_cos_cossquared}}, it is interesting to notice that the effect of the gravitational wave depends on the ratio of its amplitude with the angular frequency of the electromagnetic field, whereas the overall size of electromagnetic radiation-reaction effects depends, as in the flat spacetime, on the square of the amplitude of the electromagnetic field. In particular, close to the resonance $\lambda=1$ the gravitational wave would enhance the linear increase of radiation-reaction effects by a factor of 9/4. 

One can also study the case of a sinusoidal electromagnetic wave: $\blue{a}_i(\phi)=-\delta_i{}^1 \blue{a}_0\cos(\omega\phi)$. The function $w(\varphi)$ is given by
\begin{equation}
\label{w_2}
w(\varphi)=1+
\omega\tau_eu_0^-\xi_0^2
\int_{\varphi_0}^{\varphi} d\tilde{\varphi} \, \frac{\sin^2(\tilde{\varphi})}{\cos^2(\lambda\tilde{\varphi})}.
\end{equation}
This integral can also be computed analytically but it is more instructive to consider directly the resonant case $\lambda=1$. In this case, in fact, one obtains that radiation-reaction effects scale as $\tan(\varphi)-\varphi$ unlike linearly as in Minkowski spacetime.
Independently of a possible experimental observation, we find remarkable that a gravitational wave can significantly modify the electromagnetic radiation reaction depending on the ratio of its amplitude with the frequency of the electromagnetic wave rather than with its amplitude.

In conclusion, we have found the first exact analytical solution of the DWBH equation in the case of copropagating electromagnetic and gravitational plane wave, both featuring arbitrary frequency content and polarization. The case of a monochromatic  electromagnetic plane wave propagating along a constant gravitational plane wave in a finite phase interval has been worked out explicitly, and it has been shown how the presence of the gravitational wave can qualitatively alter electromagnetic radiation-reaction effects. We have underlined the physical importance of the found solution as it relates, via the Penrose limit, to the motion of an ultrarelativistic charge in an arbitrary combined electromagnetic and gravitational field. \blue{In this context, the present results can also be used to determine the classical electromagnetic and gravitational radiation of a charge in such background fields.}

\begin{acknowledgments}
This material is based upon work supported by the U.S. Department of Energy [National Nuclear Security Administration] University of Rochester ``National Inertial Confinement Fusion Program'' under Award Number DE-NA0004144.

This report was prepared as an account of work sponsored by an agency of the United States Government. Neither the United States Government nor any agency thereof, nor any of their employees, makes any warranty, express or implied, or assumes any legal liability or responsibility for the accuracy, completeness, or usefulness of any information, apparatus, product, or process disclosed, or represents that its use would not infringe privately owned rights. Reference herein to any specific commercial product, process, or service by trade name, trademark, manufacturer, or otherwise does not necessarily constitute or imply its endorsement, recommendation, or favoring by the United States Government or any agency thereof. The views and opinions of authors expressed herein do not necessarily state or reflect those of the United States Government or any agency thereof.

The authors gratefully acknowledge insightful discussions with S. Gralla \blue{and express their appreciation to an anonymous referee for drawing their attention to the paper \cite{Kunzle1968}.}

\end{acknowledgments}

%

    \end{document}